# Design , Simulation and Feasibility Analysis of Bifacial Solar PV System in Marine Drive Road, Cox's Bazar

Abdullah Al Mehadi, Mirza Muntasir Nishat, Fahim Faisal, Ahmed Raza Hasan Bhuiyan, Mohyeu Hussain and Md Ashraful Hoque
Department of Electrical and Electronic Engineering
Islamic University of Technology
Dhaka, Bangladesh
Email: {*abdullahmehadi, mirzamuntasir, faisaleee, ahmedraza, mohyeuhussain, mahoque* }@iut-dhaka.edu

*Abstract*—This paper proposes a design and simulation based investigative analysis of a vertically mounted bifacial solar photovoltaic model in Marine Drive Road, Cox's Bazar. Cox's bazar is a famous tourist destination which seems to be a flexible site for implementing such energy harvesting system without affecting the nearby eco-system and solves the existing land shortage problem. Moreover, the infrastructure will provide insulation to noise related problem faced by nearby residents, arising from traffic noises. A model road of 200 meters is reconnoitered for energy harvesting by solar power using three prominent software namely PVSOL, PVsyst, and SAM where a promising mean annual yield of 70492.9 kWh is obtained, and the bifacial gain is calculated to be 12.26%. In addition, a deviation analysis is performed among each of the software and it is found that PVSOL and PVsyst have shown less deviation. Furthermore, a comprehensive financial analysis shows total installation cost to be 84759.74$.

I. INTRODUCTION

Over the last several decades, the world has seen rapid industrialization which came at a cost of affecting climate and natural ecosystem. Since the pre-industrialization, the average earthly temperature is uprising which has resulted in climate change that has caused ecological imbalance and is affecting sustainable social and economic development [1]. Scientists estimate that if the current trend continues, warming of 1.5 degrees would be reached between the years 2030 and 2052 [2]. It is high time we realized the dangerous consequence of global warming and analyzed how to get prepared to face any global phenomenon that threatens the life of the entire planet [3]. Global carbon emission, acting as a primary driver of climate change and global warming is mostly caused due to power generation by conventional fossil fuel [4]. Although renewable energy stands at a promising place to supply the ever-increasing energy demand of the world, it is not being utilized by many middle and low-income countries [5]. In developing countries like Bangladesh, where population density is relatively quite high, the energy crisis is a significant concern. With limited natural resources, it is challenging to supply the enormous demand of the country through fossil fuel-based energy only. The utility electricity sector in Bangladesh has a grid with an installed capacity of 21,419 MW as of September, 2019 [6]. Although 90% of the population has access to electricity, per capita energy consumption is relatively low.

A considerable improvement is needed in the renewable energy sector as it accounts for only 3.3% of total production [7]. In-addition due to a shortage of land the country faces a huge challenge to construct a large power generation plant. This often leads to resident relocation or forest clearing to provide suitable land for power plants [8].

Due to the cost-effectiveness and eco-friendly nature, solar energy is one of the widely available carbon-free energy sources used all over the world [9]. Solar energy utilizes energy from the sun to produce electricity that can be supplied to the grid or stored in batteries. Furthermore, it does not emit anthropogenic gases into the atmosphere thus, provides a free source of clean energy [10]. Two categories of solar modules are commonly used, namely monofacial and bifacial solar modules. Bifacial PV panels showed a remarkable feat in harvesting energy from the sun. These modules can utilize the sunlight entering the module from both the front and rear sides of the panel, unlike the traditional monofacial panel [11-12]. Moreover, an increment in efficiency is observed up to 35% by using the bifacial modules, which also contributes to improve power density and reduce area requirements [13-14]. As the demand for renewable energy is increasing significantly in residential, commercial and even in tourist places like Cox's- Bazar, implementation of novel energy model is a dire need. The large-scale deployment of solar energy would have a significant effect on our living environment. Bangladesh is a densely populated country where space is scarce. A few hundred square kilometers of solar PV farms would have to be implemented in order to meet our climate objectives. The large-scale roll-out of renewable energy would have a negative impact on our living condition. This dilemma cannot be overcome without the creative implementation of the PV system. The energy output of bifacial solar panels is less dependent on position and orientation since it can capture both direct and diffuse irradiation effectively, thus provides more opportunities for multifunctional usage or higher yield per hectare.

Several researches have been done in renewable system modeling in Bangladesh. *Khan et. al* proposed a hybrid off-grid energy system that can be implemented on Sandwip Island of Bangladesh [15]. However, *Mahmud et. al* designed a solar highway model for several national roads by using bifacial PV modules [16]. Furthermore, *Kabir et. al* contributed to modeling a hybrid power plant by photovoltaic and wind as a renewable energy source and diesel source as a non-renewable stand by energy source [17]. *Islam et al.* investigated the wind characteristics and assessed wind energy potential for the Coxes-Bazar area [18]. Since, renewable energy is a viable option to meet the ever-increasing energy crisis of the world, the potential feasibility analysis of PV energy harvesting system provides an innovative approach to solve the existing energy crisis problem in a land constrained countries like Bangladesh. Although the verticall3y mounted bifacial PV system has been analyzed previously before [16], the analysis using any dedicated PV system simulation software and comparing energy yield among the software's have not been done before. In-addition, infrastructure integrated bifacial PV system acting as a noise barrier structure gives a new perspective and a noble idea for investigation. An effective method of noise prevention using photovoltaic modules was first demonstrated in A13 in Switzerland in 1989, where the installed capacity is 100 kW. The 2208 modules generate 100MWh per year [19]. After that, the solution was also applied in some other European countries. In the Netherlands, the PV noise barrier was installed along the A9 motor path with a capacity of 220 kW, which generates 176 MWh per year [20]. Thus, it is possible to build various photovoltaic noise barriers, considering the features of the roads, the construction of the barrier, the height of the barrier and other factors that affect it.

In this paper, we propose the Marine drive road as the site of our research, which is situated at the Chittagong division and extends from Cox's Bazar to Teknaf along the Bay of Bengal. The area has perceived major development after the construction of Marine drive road which has inaugurated in 2017. This has made the remote part of the district much more accessible by transport, thereby has turned out to be a famous tourist destination. The longest sea beach has witnessed a plethora of tourists in recent times which rear the demand for power generation in this coastal area to facilitate the domestic and international visitors. In this regard, the vertically mounted bifacial PV system will emerge as an efficient method of using solar energy and generating electricity, as well as providing the local community with effective sound insulation against traffic noises. However, the hardware installation of the solar panels is quite a challenge, especially in a marine environment like Cox's Bazar. Hence, software-based investigation with bifacial PV panels opens a new window to attain knowledge and ideas so that constructive decisions can be made before hardware implementation.

Therefore, our main objective is to design and develop a grid-connected PV system by utilizing vertically mounted bifacial modules to contribute in solving the problem of the energy crisis. This is primarily achieved with the help of three of the prominent PV designing software namely PVSOL, PVsyst, SAM (System Advisory Model). The software-based analysis will be able to contribute to speculating the feasibility and applicability of the system and thereby, making a proper estimation for hardware implementation. In Section II, details of the site and its weather profile are exemplified followed by the description of the solar PV model with 3D design in PVSOL. Afterward, in Section III, the results of the simulation of the model are presented and the performance of each of the software is demonstrated with deviation analysis. Hence, the work is concluded in Section IV declaring promising outcomes and immense potential.

## II. PROPOSED SYSTEM

### A. Site Details

The Cox's Bazar-Tekhnaf Marine Drive Road is approximately 80 km long comprising a magnificent view of the longest uninterrupted sea beach of the world and is situated at a height of 3m above sea level. The geographical location of the road and abundance of high solar irradiation in this vicinity has made it a suitable place for the utilization of solar energy.

TABLE I. GLOBAL IRRADIANCE AND TEMPERATURE (FROM METEONORM)

| Month | Direct Irradiance (W/m$^2$) | Diffuse Irradiance (W/m$^2$) | Temperature ($^0$C) |
|---|---|---|---|
| Jan | 150.7 | 37.1 | 21.2 |
| Feb | 154.4 | 45.0 | 23.6 |
| Mar | 184.3 | 75.1 | 26.7 |
| Apr | 189.0 | 85.1 | 28.4 |
| May | 177.0 | 89.7 | 29.1 |
| Jun | 152.7 | 90.1 | 28.2 |
| Jul | 125.7 | 90.2 | 28.2 |
| Aug | 143.8 | 91.6 | 28.4 |
| Sep | 144.3 | 70.9 | 28.2 |
| Oct | 151.2 | 73.3 | 28.3 |
| Nov | 152.6 | 38.9 | 25.7 |
| Dec | 150.1 | 28.2 | 23.0 |

The average annual solar radiation received by the coastline is around 4.77 kWh/m$^2$/day [21]. Moreover, the climate of Cox's Bazar is classified as tropical and so, rainfall is frequent in most months of the year, and the short dry season has little effect on lowering the annual rainfall.

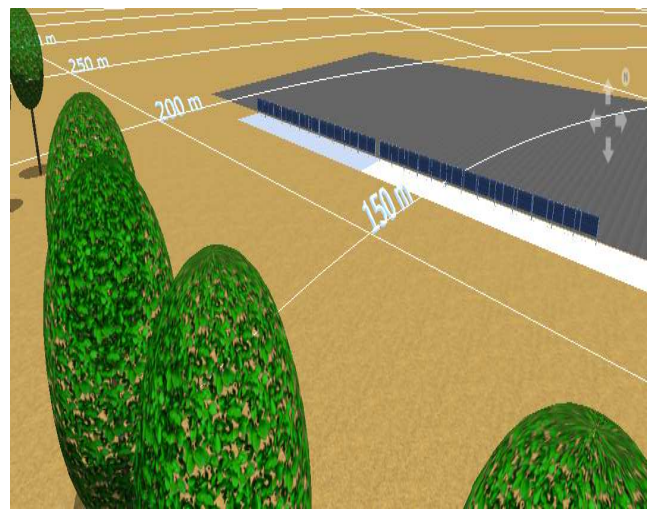

Fig. 1. 3D design of Marine drive road in PVSOL, view-1

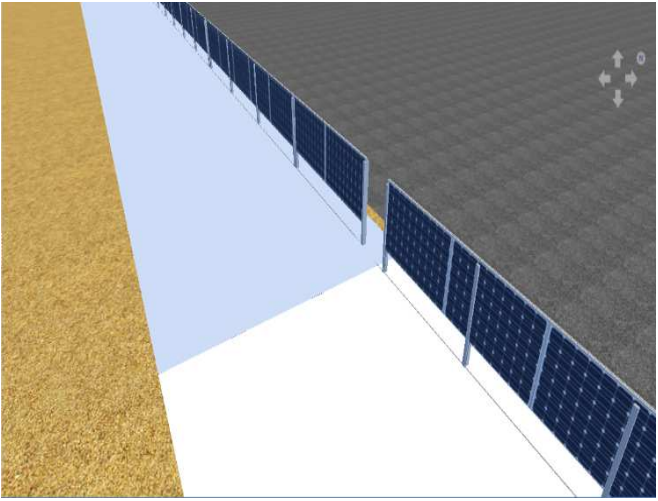

Fig. 2. 3D design of Marine drive road in PVSOL, view-2

The annual rainfall is 3770 mm (148.4 inches) [22]. However, the average annual temperature is observed to be 25.6 °C (78.2 °F). The global irradiance and temperature data are tabulated in Table I which are taken from Meteonorm.

### B. Design of the system

The design of the proposed system is accomplished by using PVSOL software (Fig. 1 and Fig. 2). A length of 200 meters is taken into consideration for primary simulation and analysis. Hence, a total number of 117 modules were placed within the specified length. The schematic diagram of the overall system is depicted in Fig. 3.

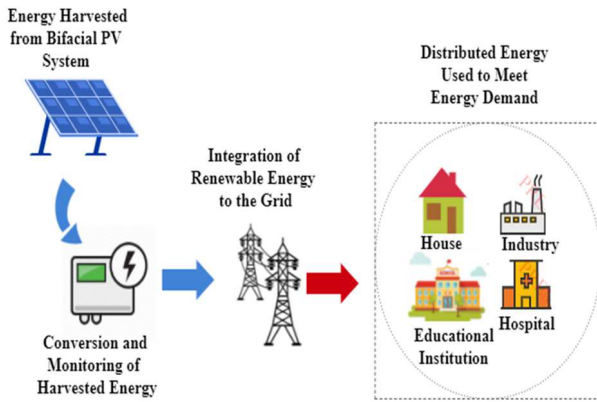

Fig. 3. Schematic diagram of the proposed system

### III. SOFTWARE SIMULATION AND RESULTS

Three popular software namely PVSOL, PVsyst, SAM have been utilized to estimate the energy generated from the proposed PV energy model. At first, the 3D model is constructed and simulation parameters are considered according to Table II. After that, the PV panels are oriented and system components are configured. Then, meticulous simulations have been performed to achieve monthly energy profiles from each of the software.

LG bifacial modules have been chosen as solar PV modules as they exhibit relatively low degradation and longer life [23]. Moreover, a total of seven inverter rated at 5000W from Huawei technologies have been selected. Thus, DC to AC ratio of 1.17 was obtained for optimum operation. Albedo was considered as 0.65 to ensure maximum reflection from the white painted surface beside the road.

TABLE II. SIMULATION PARAMETERS FOR BIFACIAL SYSTEM

| Parameters | Values | Parameters | Values |
|---|---|---|---|
| PV Model | LG340N1T-v5 | Tilt | 90° |
| Panel rating | 340 W | Azimuth | 240° |
| No. of Panels | 117 | Albedo | 0.65 |
| Total Installed Capacity | 39.78 kW | Soiling loss | 5% |
| Nominal Efficiency | 20.6 | Mismatch Loss | 2% |
| Maximum Power | 340.2 W | Diode Loss | 0.5% |
| Maximum Power Voltage | 34.4 V | Mounting Height | 2 m |
| Maximum Power Current | 9.9 A | Length | 200 m |
| Open Circuit Voltage | 40.8 V | Bifaciality | 0.7 |
| Short Circuit Current | 10.4 A | Nominal Module Operating Temperature | 42 ± 3 |
| Number of Inverters | 7 | Inverter Rating | 5000 W |
| DC to AC ratio | 1.17 | | |

TABLE III. MONTHLY ENERGY PROFILE FOR BIFACIAL PANELS

| Month | PVSOL (kWh) | PVsyst (kWh) | SAM (kWh) | Average (kWh) | Standard Deviation |
|---|---|---|---|---|---|
| Jan | 5927.3 | 5789 | 5766.59 | 5827.63 | 71.07 |
| Feb | 5654.3 | 5635 | 5846.79 | 5712.03 | 95.61 |
| Mar | 6686.7 | 6645 | 6834.33 | 6722.01 | 81.23 |
| Apr | 6715.4 | 6914 | 6481.46 | 6703.62 | 176.78 |
| May | 6420.7 | 6402 | 5966.18 | 6262.96 | 210.00 |
| Jun | 5795.3 | 5652 | 4943.63 | 5463.64 | 372.33 |
| Jul | 5384.7 | 4807 | 5394.04 | 5195.25 | 274.56 |
| Aug | 5500.7 | 5422 | 5491.65 | 5471.45 | 35.16 |
| Sept | 5673.6 | 5450 | 5468.00 | 5530.53 | 101.43 |
| Oct | 6187.4 | 5739 | 5489.03 | 5805.14 | 288.91 |
| Nov | 6104.0 | 5698 | 5963.93 | 5921.98 | 168.38 |
| Dec | 6135.3 | 5714 | 5780.68 | 5876.66 | 184.90 |
| Total | 72,185.4 | 69,867 | 69,426.31 | 70,492.9 | 1210.22 |

High albedo also ensures high diffused radiation being reflected from the surface to reach both sides of the panels [24]. For simulation, soiling loss was set at 5% only, since vertically mounted modules suffer less from soiling loss compared to Monofacial panel [25]. Moreover, the mounting surface also plays an important role in receiving light from the surrounding which results in bifacial gain [26]. Thus, mounting height was considered as 2 m above the ground for receiving maximum reflected light off the ground and the tilt angle was set at 90, azimuth was considered as 240 degrees keeping it in-line with the orientation of the road. On the other hand, mismatch loss and diode loss were set at a default value of 2% and 0.5% respectively. The simulation parameters are tabulated in Table II.

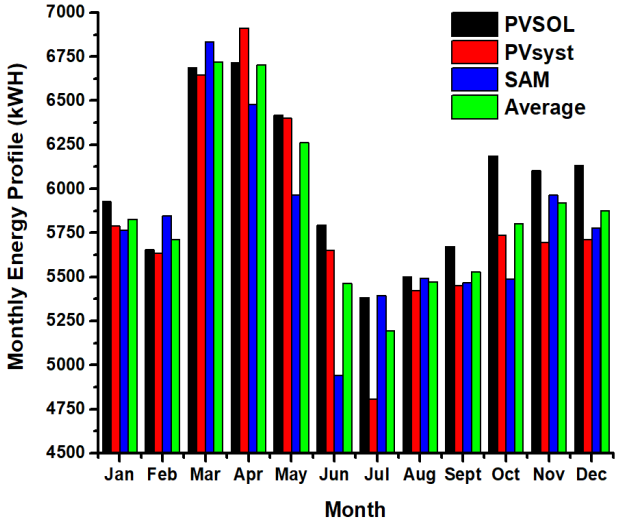

Fig. 4. Comparison of Monthly Energy Profile Among Three Software (PVSOL, PVsyst, and SAM)

The annual yield was 72185.4 kWh, 69867 kWh, 69426.31kWh in PVSOL, PVsyst, and SAM respectively. Hence the total average yield was 70492.9 kWh. Table III shows all the simulated data in each month for the three software as well as the standard deviation. The highest yield was obtained during March when both direct and diffuse irradiation was considerably high. Although April had the highest direct and diffuse irradiation value, the temperature was quite high causing the panels to lose efficiency. The yield falls gradually after April until the lowest yield was obtained in July due to rain and cloudy weather. The mean yield starts to increase from October which can be linked with increase in global irradiation and decrease in temperature. The graphical representation of the monthly energy profile of each of the software is illustrated in Fig. 4.

IV. DISCUSSION

Designing solar PV models in the marine environment emanates costs as this is frequently exposed to salty water due to winds and cyclones which make the panels vulnerable to rust. Moreover, salt deposit on the surface of the panel makes light propagation difficult [27]. Furthermore, the strong wind might cause the mounting system to break. These entire factors bring a negative consequence in annual yield and efficiency. Nevertheless, the average annual yield was found to be 70492.9 kWh for 200-meter road. If solar modules are planned for the entire length of the road (80 Km) total energy generated will be approximately 28,197,160 kWh which will carry an immensely positive impact on the energy generation of this coastal area. To estimate the accuracy of the simulation, deviation analysis was accomplished to investigate similarities and differences in algorithm and simulation method among each of the software which is presented below:

A. Deviation Analysis

The monthly deviation analysis among each of the PV simulation software is tabulated in Table IV. The graphical representation is depicted in Fig 5. There seems to be a large energy difference between PVSOL versus SAM and PVsyst versus SAM. The monthly energy difference was quite high in this case because SAM takes a few more losses into account. On the other hand, a small difference is seen between PVSOL and PVsyst as similar energy estimation algorithms are used in both software. Furthermore, there is a slight difference in the mathematical model used in each of the software and how irradiation value and yield are calculated.

TABLE IV. DEVIATION ANALYSIS

| Month | PVSOL and PVsyst (kWh) | PVSOL and SAM (kWh) | PVsyst and SAM (kWh) |
|---|---|---|---|
| Jan | 138.3 | 160.71 | 22.41 |
| Feb | 19.3 | -192.49 | -211.79 |
| Mar | 41.7 | -147.63 | -189.33 |
| Apr | -198.6 | 233.94 | 432.54 |
| May | 18.7 | 454.52 | 435.82 |
| Jun | 143.3 | 851.67 | 708.37 |
| Jul | 577.7 | -9.34 | -587.04 |
| Aug | 78.7 | 9.05 | -69.65 |
| Sep | 223.6 | 205.6 | -18 |
| Oct | 448.4 | 698.37 | 249.97 |
| Nov | 406 | 140.07 | -265.93 |
| Dec | 421.3 | 354.62 | -66.68 |

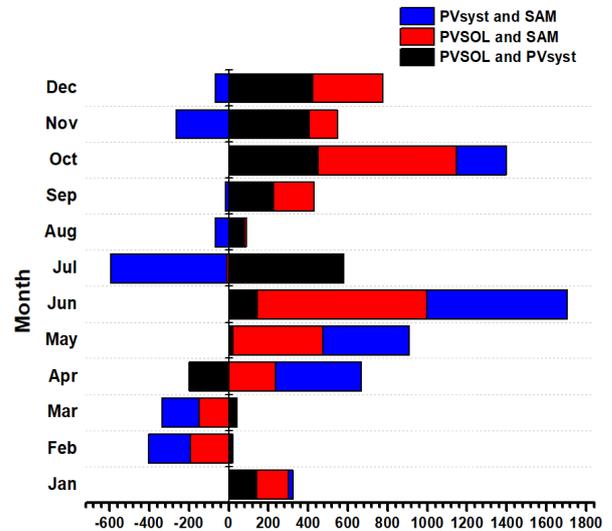

Fig. 5 Yield deviation among PVSOL, PVsyst, and SAM

The weather profiles used by each of the software accounts for the difference too, since the profiles are obtained from various sources depending on the software. The standard deviation shown in Table-III depicts a clear picture of the gap between the mean and obtained monthly result in each of the software. There is a large deviation seen during June, July and October, with the highest deviation of 372.33 seen on June. On the other hand, low deviation is seen during the month of January, February, March and August, with the lowest deviation of 35.16 is seen on August.

*B. Financial Analysis*

A rough estimation of the proposed project is illustrated for 200-meter model road where it is observed that the total installation cost becomes 84759.74$ approximately. The price of the solar panels is considered as 261.57$ [28]. The price of one inverter is $1495.41 [29]. The cost of the PV panel integrated noise barrier system is assumed to be 1$/W and wiring and other expenses such as fuses, switch gear and relays are considered to be 2378$. Based on total installation cost, the per unit energy cost is found to be 1.20$/kWh. Table V presents the financial analysis of the system.

TABLE V. FINANCIAL ANALYSIS

| Description | Unit price ($) | Quantity | Price ($) |
|---|---|---|---|
| LG PV modules | 261..57 | 117 | 30603.69 |
| Mounting system | 1$/W | 117 | 39780.00 |
| Inverter | 1495.41 | 7 | 10467.87 |
| Wiring and miscellaneous cost | | | 2378.00 |
| ATV (5% of solar panel cost) | | | 1530.18 |
| Total Installation Cost/Capital Cost | | | 84759.74 |
| Maintenance and miscellaneous cost (%5 of total capital cost) | | | 4237.987 |

*C. Bifacial Gain*

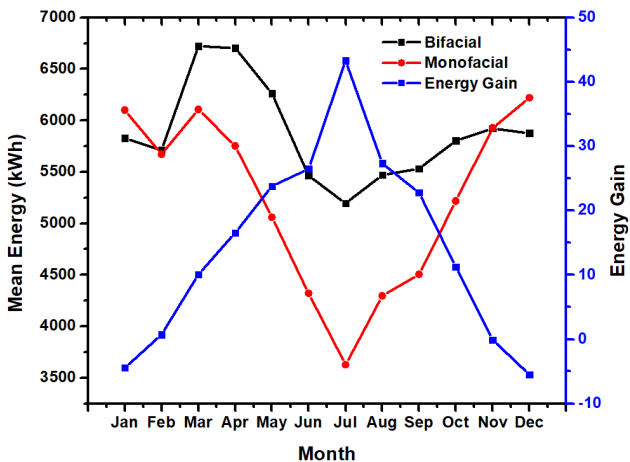

Fig. 6. Graphical representation of mean energy (kWh) of monofacial system, bifacial system and bifacial energy gain

To analyze the prospect of bifacial system compared with traditional monofacial panel, a comparison analysis has been done to find the bifacial gain of the system. Monofacial system were set at 180-degree azimuth and tilted at 24 degrees for optimum setup. Necessary simulations were performed for monofacial system where the mean annual energy is obtained 62.8 MWh. Thus, annual bifacial gain is found to be 12.26%. Table VI shows the mean monthly energy production and bifacial gain between two systems. This is graphically represented in Fig.6.

TABLE VI. COMPARATIVE ANALYSIS OF BIFACIAL AND MONOFACIAL

| Month | Mean Energy of Bifacial System (kWh) | Mean Energy of Monofacial System (kWh) | Energy Gain b (%) |
|---|---|---|---|
| Jan | 5827.63 | 6101 | -4.48 |
| Feb | 5712.03 | 5672 | 0.70 |
| Mar | 6722.01 | 6107 | 10.07 |
| Apr | 6703.62 | 5752 | 16.54 |
| May | 6262.96 | 5060 | 23.77 |
| Jun | 5463.64 | 4320 | 26.47 |
| Jul | 5195.25 | 3626 | 43.27 |
| Aug | 5471.45 | 4297 | 27.33 |
| Sep | 5530.53 | 4505 | 22.76 |
| Oct | 5805.14 | 5217 | 11.27 |
| Nov | 5921.98 | 5928 | -0.10 |
| Dec | 5876.66 | 6220 | -5.51 |
| **Total** | **70492.9** | **62805** | **12.26** |

V. CONCLUSION

In this paper, a design, simulation, and analysis of a bifacial solar PV panel integrated noise barrier infrastructure along the Marine drive road have been presented. Using three different types of software, a detailed 3D design was formed and the annual performance of the solar panels was simulated. The proposed design showed a promising yield of 70492.9 kWh annually by utilizing 117 modules rated 340W which is mounted beside a 200-meter-long road. If this is extrapolated for the entire road (almost 80 Km), the annual generated yield will be almost 28,197,160 kWh (approximately) while at the same time providing significant blockages to traffic noise after integrating with noise barrier infrastructure. Furthermore, an approximate financial analysis revealed the total installation cost to be 84759.74$. In-addition, a deviation analysis between 3 software showed less deviation between PVSOL and PVsyst while the comparison between bifacial and monofacial panel-based system showed an energy gain (Bifacial gain) of 12.26%. Hence, this clean energy initiative will prevent carbon emission that would otherwise be released from a conventional fossil fuel-based power plant. So, this promising source of green energy can be implemented in a popular tourist destination like Cox's Bazar with a view to turning it into an eco-friendly city.